\documentclass[reprint, twocolumn, nofootinbib, amsmath, amssymb, aps, pra, floatfix]{revtex4-1}

\usepackage[english]{babel}
\usepackage[utf8x]{inputenc}
\usepackage[T1]{fontenc}
\usepackage{graphicx}
\usepackage{dcolumn}
\usepackage{bm}

\makeatletter

\newcommand{\Rb}{$^{87}$Rb}
\newcommand{\ketto}[2]{\ket{#1}\to\ket{#2}}
\newcommand{\sub}[1]{_{\mathrm{#1}}}

\usepackage[colorinlistoftodos]{todonotes}
\usepackage[colorlinks=true, allcolors=blue]{hyperref}
\usepackage{eqnarray,tabstackengine}
\usepackage{physics}

\definecolor{green2}{rgb}{0.0, 0.5, 0.0}

\begin{document}

\title{Multi-Frequency Coherence Control of Radio-Frequency-Dressed States}

\author{B.~Foxon}
\author{S.~Jammi}
\author{T.~Fernholz}
\affiliation{School of Physics \& Astronomy, University of Nottingham, University Park, Nottingham NG7 2RD, United Kingdom}

\begin{abstract}
We demonstrate engineering of a narrow microwave transition between trappable states in radio-frequency-dressed $^{87}$rubidium, reducing the static field dependence. A single-frequency, off-resonant microwave field allows for the suppression of the differential Zeeman shift arising from the nuclear magnetic moment to at least first order. The field dependence can be suppressed further with additional dressing fields, which we demonstrate experimentally with two microwave frequencies. The engineered transition can thus be used in a range of cold atom schemes that rely on coherent state superpositions.
\end{abstract}

\maketitle

\section{Introduction}

Radio-frequency dressing of magnetically trapped alkali atoms can provide a versatile basis for the development of complex potential landscapes. Through the combination of static and oscillating magnetic fields, geometries with interesting topologies such as double-wells \cite{Schumm2005, harte}, lattices~\cite{courteille}, ring-shaped potentials~\cite{perrin, sherlock2011, vKlitzing}, hollow shells from spheres and ellipsoids~\cite{tononi} to toroidal surfaces~\cite{fernholz, gentile} are enabled with a range of applications \cite{dubessy2025}. A high degree of versatility arises from position and polarisation-dependent coupling strength in in-homogeneous fields by driving resonances at the atomic Larmor frequency. This can, for example, lead to sub-wavelength structure in optical potentials~\cite{lundblad, baranov} and to state-dependent trapping and transport~\cite{navez, atkocius}. Off-resonant dressing at radio-frequencies (coupling Zeeman sublevels) and microwave frequencies (coupling hyperfine levels) has been used to modify the atomic level structure to change the effective Land\'e-factor~\cite{beaufils}, and alter field dependencies to create decoherence-free sub-spaces~\cite{sinuco_2} and introduce continuous dynamical noise decoupling to stabilize an atomic clock frequency \cite{pelzer}.
In the context of magnetically trappable clock states~\cite{treutlein, szmuk}, radio-frequency (RF) and microwave (mw) dressing have been investigated to reduce field sensitivity to higher orders near magic field conditions~\cite{schumm, fortagh}. 

In this paper, we investigate multi-frequency microwave dressing in order to modify state-dependent radio-frequency dressed potentials. Our particular interest is in controlling a narrow transition to generate coherent superpositions of two trappable clock states that can be independently controlled and used in the context of trapped atom interferometry without free propagation~\cite{stevenson, johnson, ammar}, recently termed tractor atom interferometry~\cite{duspayev}. This is somewhat different from compensatory dressing of bare states because field sensitivity does not only arise directly from differential Zeeman shifts but from state-dependent resonance conditions for the primary RF dressing. This issue has been addressed previously using bichromatic, two-field dressing~\cite{mas}, which comes with limitations in trapping geometries as the two frequencies should remain separate across a spatially varying, static field-dependent decomposition into orthogonal polarization components of the RF fields. The method of secondary dressing explored here is a suitable alternative for the scenario considered in~\cite{stevenson} and may find  
application in cancelling residual potential variations in gravity-compensated~\cite{guo} or microgravity bubble traps~\cite{carollo} to study homogeneous quantum gases in new topologies. 

We theoretically investigate the static magnetic field dependence of the ground state of RF-dressed alkali atoms when secondary mw-dressing is applied. We experimentally demonstrate engineering of the trappable clock transition in RF-dressed $^{87}$rubidium with a significant reduction of magnetic field sensitivity using single and two-colour microwave fields. 

\vspace{-2mm}

\section{Radio-Frequency Dressed Hyperfine Transitions}

\begin{figure*}[thb]
    \centering
    \includegraphics[scale=0.47]{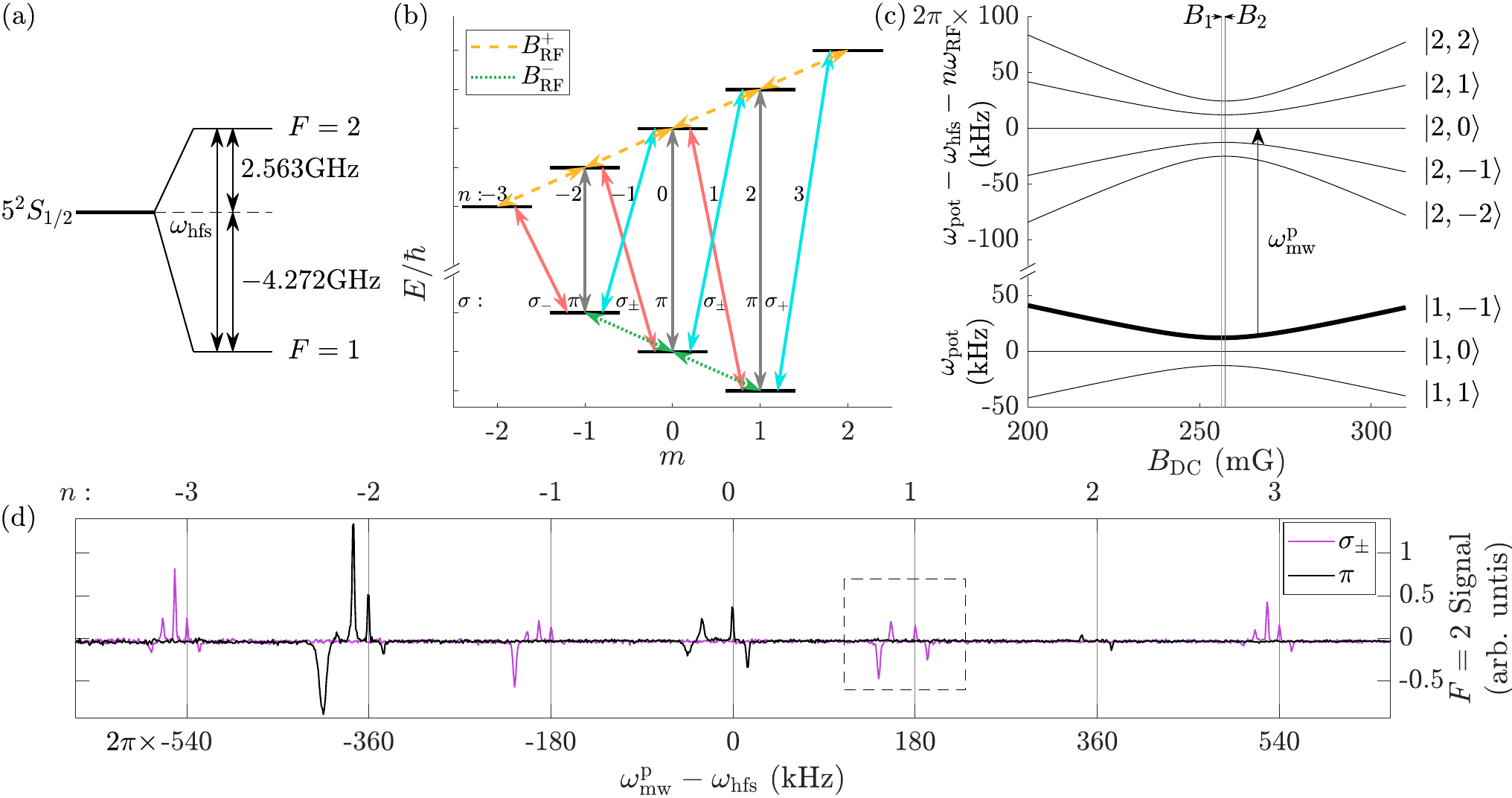}
    \caption{Microwave (mw) spectroscopy of radio-frequency (RF)-dressed \Rb. Two ground-state hyperfine levels with spin magnitudes $F=1$ and $F=2$ as shown in (a) are separated by an energy $\hbar \omega_\text{hfs}=h\times6.835~\text{GHz}$ ~\cite{steck}. The weak-field splitting into non-degenerate, magnetic sublevels $\ket{F,m}$ shown in (b) permits nine polarisation-dependent mw transitions between hyperfine manifolds, with two pairs of nearly degenerate frequencies. 
    Neighbouring states are coupled with AC magnetic fields in the RF range (here at $\omega_\mathrm{RF}=180\times2\pi~\text{kHz}$), with opposite $\sigma_\pm$ polarisations due to opposite signs of $g$-factors $g_F$ for the two manifolds.
    This dressing results in a static-field-dependent spectrum of quasi-energies shown in (c) with avoided level crossings where the Larmor precession frequency of the spin $\mathbf{F}$ is resonant with the RF. The resonant field values $B_F$, marked with vertical lines, differ due to the magnetic moment of the atomic nucleus.
    The RF dressing leads to a rich microwave spectrum, where seven spectroscopic groups labeled $n=-3$ to $3$ emerge, each corresponding to multi-photon processes involving one of the bare mw transition frequencies. Experimental data for atoms prepared in a pure, dressed state $\ket{1,-1}$ are shown in (d) for probing at frequency $\omega_\mathrm{mw}^\mathrm{p}$ with two orthogonal, linear polarisations. Each group reflects five dressed target states in the upper ($F=2$) level. The highlighted region of $n=1$ transitions is examined in more detail in Fig.~\ref{fig:2}.}
    \label{fig:1}
\end{figure*}

We first revisit the main aspects of our previous analysis of mw spectroscopy of RF-dressed atoms~\cite{sinuco}. We begin by establishing the energy level structure of an RF-dressed alkali atom in the weak field regime, in the ground state hyperfine levels $F=1$ and $2$ of \Rb~(Fig.~\ref{fig:1}(a) and b). The RF drives transitions $\ket{F,m}\leftrightarrow\ket{F,m\pm1}$, which are marked in yellow and green in Fig.~\ref{fig:1}(b) for $\sigma^+$ and $\sigma^-$ polarisations, respectively. If we define the frequency $\omega\sub{RF}$ to be positive, this sets the resonant static field as $B_F=\hbar\omega\sub{RF}/(\mu_B|g_F|)$, where $g_F$ is the level-dependent Land\'e $g$-factor.
Neglecting oscillating terms in the rotating frame, the coupling Hamiltonian reduces to the interaction of the atomic spin with an effective field according to 
\begin{equation}
    H\sub{eff}=\frac{\mu_Bg_F}{\hbar}\boldsymbol{F}\cdot\underbrace{\left(\frac{B\sub{RF}^\pm}{2}\boldsymbol{\mathrm{e}_x}+(B\sub{DC}-B_F)\boldsymbol{\mathrm{e}_z} \right)}_{=\boldsymbol{B}_{\mathrm{eff},F}},
    \label{Eq:Heff}
\end{equation}
where our coordinates are chosen such that $B\sub{DC}\boldsymbol{\mathrm{e}_z}$ is the static field and $B\sub{RF}^\pm$ are the amplitudes of the $\sigma^\pm$-polarized components of the RF-field.
This will lead to a set of quasi-energy levels with a dressed potential contribution of
\begin{equation}
    E_{F,\bar{m}}=\bar{m} \mu_B g_F |\boldsymbol{B}_{\mathrm{eff},F}|,
    \label{Eq:levelEnergy}
\end{equation}
with a new quantum number $\bar{m}$, which determines the projection of the spin $\boldsymbol{F}$ in the direction of the effective field.  Thus the dressing changes the static field dependence from a linear Zeeman shift into the set of dressed potentials plotted in Fig~\ref{fig:1}(c), where the RF-resonance manifests as a set of avoided crossings. 

Controllable state-dependence of these potentials arises from the fact that magnetic sublevels in the two hyperfine manifolds are coupled by orthogonal RF polarization components, i.e.\ by amplitude $B_\text{RF}^+$ for positive $g$-factor ($F=2$) and $B_\text{RF}^-$ for negative $g$-factor ($F=1$).\footnote{Alternatively, the resonance frequency can be defined as including the sign of $\omega\sub{RF}=\mu_B g_FB_F/\hbar$ with coupling of $B_\text{RF}^+$ only.}
Robustness against external magnetic field noise as well as the ability to overlap trapping potentials are required to create coherent superpositions of clock states. This condition can be fulfilled for states with identical $\bar{m}g_F$. Fortunately the g-factors of the electronic ground state hyperfine levels have nearly identical magnitude but opposite sign, because the magnetic moment is dominated by the spin of the valence electron, while the moment of inertia is dominated by the nuclear spin, thus changing mostly the sign upon a nuclear spin flip. However, the weak influence of the nuclear magnetic moment on the magnitude of $g_F$ cannot be neglected, particularly through its influence on the resonant field $B_F$, which is the issue we intend to address.

Now we consider the effect of mw-coupling between such levels, where multi-photon transitions allow coupling between any pair of magnetic sublevels from the two hyperfine manifolds.
We note the emergence of groups of transitions between hyperfine levels centred on multiples of the RF-dressing frequency $n\omega\sub{RF}$, where $n$ is the group number.  Each group of dressed transitions is linked to the bare mw transitions, which must obey the standard angular momentum selection rules. As illustrated in Fig.~\ref{fig:1}(b), $\sigma_\pm$-polarised mw may be used to probe odd numbered groups, and $\pi$-polarised mw may probe even numbered groups. However, within each of these groups any arbitrary transition between sublevels in different hyperfine manifolds may occur, except at specific static fields where couplings may become suppressed. 
The energy of a transition between dressed manifolds $\ket{F,\bar{m}}\to\ket{F+1,\bar{m}'}$ will be given by
\begin{equation}
    E_{F,\bar{m}\to F+1,\bar{m}'}^n=\hbar\omega\sub{hfs}+n\hbar\omega\sub{RF}+E_{F+1,\bar{m}'}-E_{F,\bar{m}},
    \label{Eq:unshiftedEnergy}
\end{equation}
where $\omega\sub{hfs}$ is the hyperfine splitting frequency. 

\begin{figure*}[thb]
\centering
\includegraphics[scale=0.48]{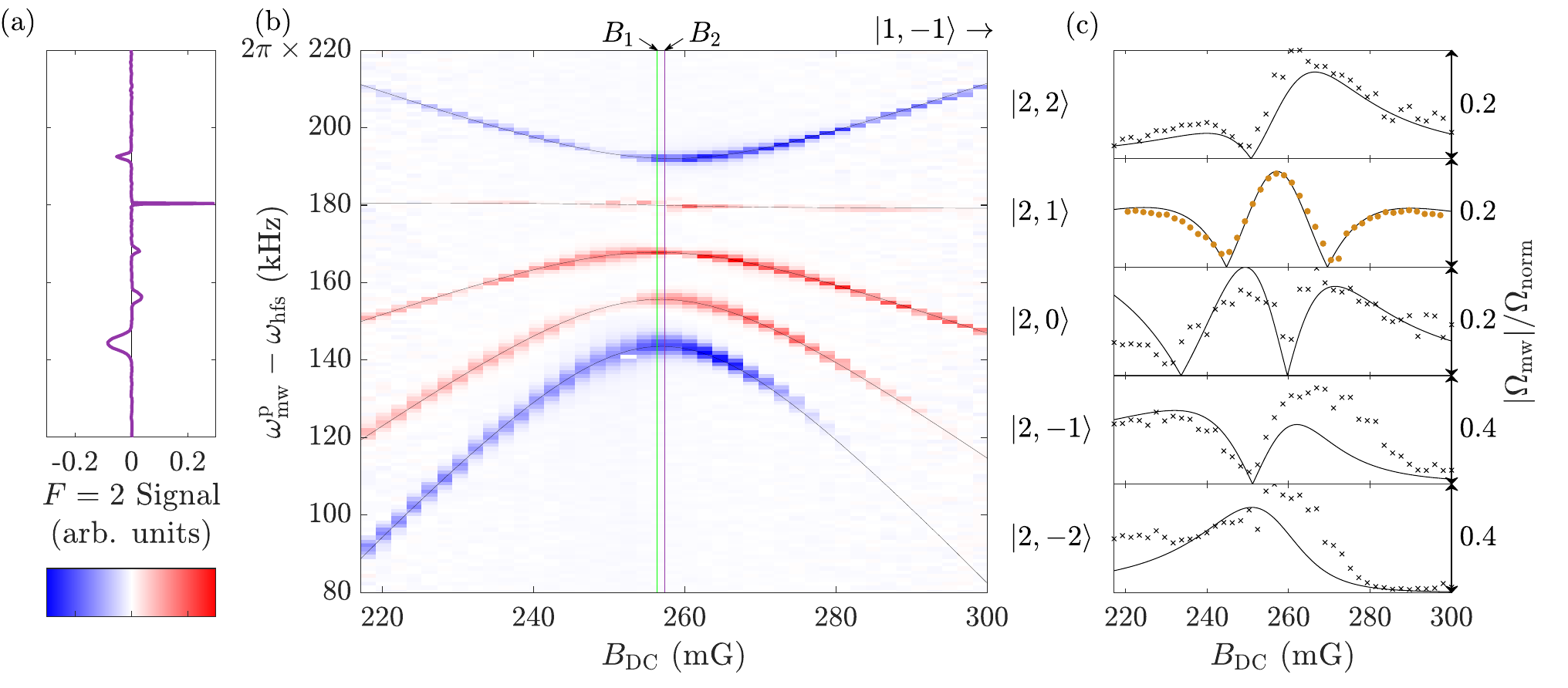}
    \caption{Experimental data for transitions within group $n=1$. We prepare atoms in the dressed state $\ket{1,-1}$ and scan the frequency $\omega_\mathrm{mw}^\mathrm{p}$ of a low-power ($B_\mathrm{mw}^\mathrm{p}=0.4$~mG) mw-probe across group $n=1$. The sample spectrum plotted in (a) has been taken near the avoided crossing and shows a birefringence signal of the final population in $F=2$ (negative for $\bar{m}=\pm2$) ~\cite{jammi}, revealing five possible transitions $\ketto{1,-1}{2,\bar{m}'}$. The dependence of this spectrum as a function of static magnetic field is shown in (b) for a somewhat higher mw-probe power ($B_\mathrm{mw}^\mathrm{p}=2$~mG). The field values for the RF-resonance in each hyperfine manifold $B_F$ are indicated by vertical lines. A simultaneous fit (solid lines) to all transition frequencies using Eqs.~\ref{Eq:Heff}-\ref{Eq:unshiftedEnergy} yields two nearly identical RF amplitudes of $B\sub{RF}^\pm= 24.5\pm0.1$~mG. The normalised coupling strengths of the transitions as a function of static field are shown in (c) together with predicted strengths from Eqs.~\ref{eq:Rabi}-\ref{eq:Rabinorm} (solid lines). Data points (black crosses) are expressed as relative mw Rabi-frequencies $|\Omega_\mathrm{mw}|/\Omega_\mathrm{norm}$, which were approximated from the peak areas in (b), measured after a $3.5~\text{ms}$ probe pulse, for all (partially incoherent) transitions except $\ketto{1,-1}{2,1}$, for which we observe coherent Rabi cycles and can measure the Rabi frequency directly (filled orange circles).}
\label{fig:2}
\end{figure*} 

In the considered weak-field regime, the coupling coefficients between bare hyperfine levels are independent of static magnetic field; the dressed picture requires a more complex formulation, however.
As shown in our previous work~\cite{sinuco}, the coupling coefficient for two dressed sublevels occurring via the bare mw-transition $\ket{F,m}\to\ket{F+1,m+\sigma}$ is given by
\begin{equation}
\begin{aligned}
    \bra{F+1,\bar{m}'}\hat{\mathrm{H}}\sub{mw}^{\sigma,n=2m+\sigma}\ket{F,\bar{m}}=&\alpha_\sigma B\sub{mw}^{\sigma} d^{F+1}_{\bar{m}',m+\sigma}(-\theta_{F+1}) \\ &\times d^F_{m,\bar{m}}(\theta_F),
    \end{aligned}
    \label{eq_couplings}
\end{equation}
where $\sigma=\pm1, 0$ for $\sigma_\pm,\pi$ polarisations. This applies to the cases of $n= 2m+\sigma$, while the coupling element is $0$ in all other cases as this would violate the standard selection rules for the bare states. $\alpha_\sigma$ is the polarisation and atomic species-dependent magnetic dipole element
\begin{equation}
\begin{aligned}
\begin{aligned}
    \alpha_\sigma=&\eta_\sigma g_J\mu_B(-1)^{F+1-m-\sigma}\\ & \times \sqrt{\frac{2I(I+1)}{2I+1}}
        \begin{pmatrix}
        F+1 & 1 & F \\ -m-\sigma & \sigma & m
    \end{pmatrix},
    \end{aligned}
    \end{aligned}
\end{equation}
where the final term is the Wigner $3j$-symbol and $\eta_\sigma=1,\pm \frac{1}{\sqrt{2}}$ for $\sigma=0,\mp1$ respectively.

The dependence of the coupling coefficients on static magnetic field is introduced via the Wigner $d$-matrices
\begin{equation}
    d^F_{m',m}(\theta_F)=\bra{F,m'}e^{-i\theta_F\hat{F}_y}\ket{F,m},
\end{equation}
due to the frequency-dependent mixing angles $\theta_F$. These describe rotations of the effective field $\boldsymbol{B}_{\mathrm{eff},F}$ for each hyperfine manifold in the rotating frame's $x,z$-plane, and are given by
\begin{equation}
    \theta_F=\frac{\pi}{2}-\mathrm{tan}^{-1}\left(\frac{B\sub{DC}-B_F}{\sqrt{2}B\sub{RF}^\pm}\right).
\end{equation}
Hence the (real, positive) Rabi frequency 
\begin{equation}
    \Omega_\mathrm{mw}^{\sigma, n} = |\hat{\mathrm{H}}_\mathrm{mw}^{\sigma, n}|/\hbar \label{eq:Rabi}
\end{equation}
is likewise a function of $B_\mathrm{DC}$. The locations of points with vanishing coupling depend on the nuclear magnetic moment, as well as the resonant static field $B_F$ through the mixing angle.

In order to validate the theory given above, we provide experimental measurements of specific transition frequencies and coupling coefficients. We use cold atoms released from a magneto-optical trap, which we prepare and analyse in free-fall. We prepare atoms in a pure $\ket{F=1,m=-1}$ state, using a sequence of mw and optical pulses (for more detail see~\cite{sinuco}). The ensemble is adiabatically converted to a dressed state $\ket{F=1,\bar{m}=1}$ by ramping up the RF-dressing field amplitude while tuning the Larmor precession frequency into resonance via the static magnetic field. A mw-spectroscopy pulse is then applied to probe atoms at a chosen frequency $\omega_\mathrm{mw}^\mathrm{p}$. We use a dispersive, optical measurement technique, where the sign and amplitude of the signal depend on $F$ and the dressed value $\bar{m}$, allowing for selective detection of populations in $F=1$ or $2$ (see~\cite{jammi}). 

Experimental mw spectra are shown in Fig~\ref{fig:1}(d) for an RF-dressing frequency of $\omega\sub{RF}=180\times2\pi$~kHz and amplitudes $B\sub{RF}^\pm\approx25$~mG with the static field tuned to the average value of RF resonance $B\sub{res}=(B_1+B_2)/2$, using $0.4$~ms mw-probe pulses. In order to detect both $\sigma_\pm$ and $\pi$-transitions, two scans are performed with different relative angles between static and linearly polarised mw-field. Seven spectral groups appear and transitions within each group are separated by the RF Rabi frequency $\Omega\sub{RF}^\pm\approx12\times2\pi$~kHz, where $\hbar\Omega\sub{RF}^\pm=\mu_B|g_F|B\sub{RF}^\pm/\sqrt{2}$.
The data was taken using relatively strong mw-probe amplitudes ($B\sub{mw}^\mathrm{p}=2.9$~mG) in order to detect most transitions easily. One should note, however, that power broadening, the occurrence of Rabi cycles, and sublevel-dependent signal strength and sign make the detected peaks unrepresentative of relative linewidths and coupling strengths. 

A low mw-power ($B\sub{mw}^\mathrm{p}=0.4$~mG) scan of the $n=1$ group shown in Fig.~\ref{fig:2} (a) reveals that one of the transitions is sharp, as the potentials of the involved states have near-identical dependence on the fluctuating and/or spatially varying static field. We scan the static field dependence of this group at higher mw-power ($B\sub{mw}^\mathrm{p}=2$~mG), and the resulting set of spectra shown in Fig.~\ref{fig:2}(b) reflects the dressed level structure seen in Fig.~\ref{fig:1}(c), as the transition energy is given by the  $\ket{2,\bar{m}'}$ potentials after subtraction of the $\ket{1,\bar{m}=-1}$ initial state potential. We use the measured field dependence of each of the transitions and a fit using Eqs.~\ref{Eq:Heff}-\ref{Eq:unshiftedEnergy} to obtain a measured RF-field strength of $B_\mathrm{RF}^\pm=24.5\pm0.1$~mG. The sharpness of the transition $\ketto{1,-1}{2,1}$ seen in (a) is clearly linked to the vanishing field dependence seen in (b).

Beyond frequency determinations, we also measure the variations of coupling strengths across the RF resonance, which can be estimated from the area under the spectroscopic peaks. The results, as extracted from Gaussian fits, are shown in Fig.~\ref{fig:2}(c) as a function of static field for four of the five transitions shown in (b). The heights are rescaled by the respective detection strength for each state, which is inherent to our detection method~\cite{jammi}, and then normalised by a single factor to account for the arbitrary overall scale of the signal. The theoretical predictions from Eq.~\ref{eq_couplings} are also plotted, scaled by
\begin{equation}
    \Omega\sub{norm}^\sigma=\frac{1}{16}\sqrt{\frac{3}{2}}|\eta_\sigma|\mu_Bg_JB^\sigma\sub{mw},\label{eq:Rabinorm}
\end{equation}
and show broad agreement in the locations of the maxima and zero crossings, even though the nature of this method precludes more accurate measurement. Another method is used for measuring the coupling coefficient of the narrow transition $\ketto{1,-1}{2,1}$, which can be determined by driving coherent Rabi cycles. The measured behaviour of Rabi frequency shows good agreement with the theoretical prediction. This sharp line is the focus of our further discussion.

 \begin{figure}[t]
    \centering
    \includegraphics[scale=0.5]{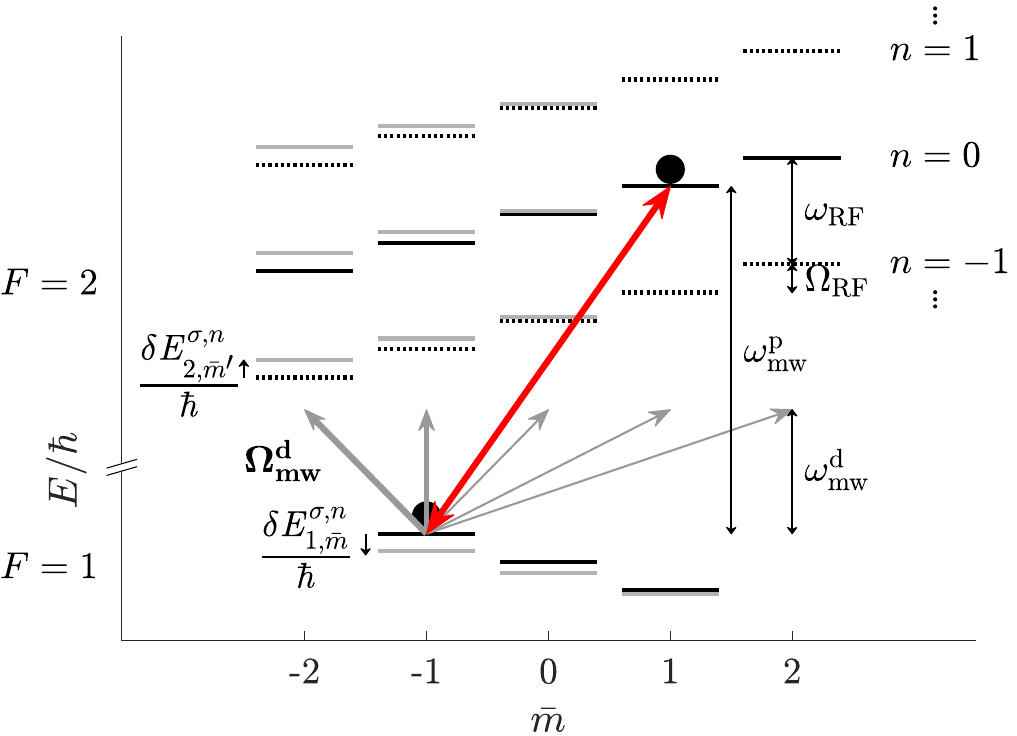}
    \caption{Schematic of RF-dressed quasi-energies with additional mw-dressing. At RF-resonance, the Rabi frequency or coupling strength $\Omega_\mathrm{RF}$ of RF-dressing determines the splitting of magnetic sublevels. The dressing frequency $\omega_\mathrm{RF}$ sets the spacing between spectral groups, which can be viewed as mw modulation sidebands arising in the rotating frame, represented here as a set of (dotted) copies of the dressed $F=2$ levels. The marked mw-probe frequency $\omega_\mathrm{mw}^\mathrm{p}$ (red), here shown within group $n=0$, is used to transfer population between the targeted clock states. An off-resonant, mw dressing field (grey) at frequency $\omega_\mathrm{mw}^\mathrm{d}$ introduces AC Zeeman shifts $\delta E$ (Eq.~\ref{Eq:deltaE}), depending on coupling strengths $\Omega_\mathrm{mw}^\mathrm{d}$. This example illustrates red-detuned dressing near group $n=-1$ with a dominant influence on the state pair $\ket{1,-1}$ and $\ket{2,-2}$, but also other levels experience energy shifts, depending on coupling strengths and detunings of all possible transitions.
    }
    \label{fig:3}
\end{figure}

\vspace{-2mm}

\section{A Trappable Clock Transition}

We use the term clock states for pairs of energy levels that have a matching response to magnetic fields. The transition frequency between two such states is independent of the external influence, and hence is robust against broadening and instability arising from field fluctuations and inhomogeneities. The Zeeman substates with $\bar{m}=0$ provide such a clock transition in the weak field regime. However as these states are individually field-independent to first order they are not magnetically trappable. In contrast, the bare or dressed states $\ket{1,-1}$ and $\ket{2,1}$ are trappable, and experience near identical potentials. There is however a small mismatch in the Land\'e $g_F$-factors for different hyperfine levels. The bare state potential difference becomes independent of the static field only in the vicinity of a field of $B\approx3.23$~G, where the second order Zeeman shift leads to equal magnetic moments. This situation has been used to demonstrate a stable atom chip-based clock~\cite{szmuk}.

In RF-dressed potentials, the situation is more complex, because the potential depends on the RF-amplitude as well as the resonance condition between bare Zeeman sublevels, which differs slightly between the hyperfine manifolds. This translates to a residual dependence of the transition frequency on the static field strength. Our approach to generating a pair of trappable clock states is to tune this static field dependence by using additional microwave fields.
The principal idea is depicted in Fig.~\ref{fig:3}, which shows quasi-energies of RF-dressed levels in $^{87}$rubidium. The two target states have a very similar field dependence, which we want to make identical by (predominantly) coupling one of the states to additional levels by a mw-field of frequency $\omega\sub{mw}^\mathrm{d}$ and using the energy shifts that arise from detuned coupling, analogously to the AC-Stark effect for electric fields. 
The energy difference between the two states $\ket{F,\bar{m}}$ and $\ket{F+1,\bar{m}'}$ will then have two additional contributions compared to
Eq.~\ref{Eq:unshiftedEnergy} of the form
\begin{equation}
    E=E_{F,\bar{m}\to F',\bar{m}'}^n-\delta E_{F,\bar{m}}+\delta E_{F',\bar{m}'}.
    \label{Eq:dressedmodel}
\end{equation}

To calculate the total shifts of the two levels of interest, we go beyond second-order perturbations of the form $\delta E=\Omega^2/\Delta$  by using an exact expression for quasi-energy shifts in driven two-level systems, but ignore changes to eigenstates. Each shift is then determined by summing over all possible transitions according to
\begin{equation}
    \delta E_{F,\bar{m}} = (-1)^p\sum_{k} \frac{\hbar\Delta_{k}}{2}\left (\sqrt{1+\frac{\Omega_{k}^2(\Delta_k)}{\Delta_k^2}}-1\right),
    \label{Eq:sumDress}
\end{equation}
with signature $p={F-I+1/2}$ and detunings 
\begin{equation}
    \hbar\Delta_k=\hbar\omega_\mathrm{mw}-|E_k-E_{F,\bar{m}}|,
    \label{Eq:deltaE}
\end{equation}
where $k$ runs over transitions from state $E_{F,\bar{m}}$ to target states with energy $E_k$ in the other hyperfine manifold, with corresponding, detuning-dependent Rabi frequencies $\Omega_k(\Delta_k)$. 

The dependence on the static magnetic field enters via both the Rabi frequency and the detuning. It is important to note that the detunings are approximately even functions of the static field deviation $B\sub{DC}-B_\mathrm{res}$, and so it is the field dependence of the Rabi frequencies that allows for the introduction of odd-order shifts near the dressed trapping minimum.
The depiction of the scheme in Fig.~\ref{fig:3} illustrates the effect of red-detuned mw-dressing near the RF-dressed transition $\ketto{1,-1}{2,-2}$, suggesting that the dominant shifts occur in these two states, therefore increasing the targeted transition frequency via the shift of $\ket{1,-1}$. However the microwave dressing also leads to small shifts on neighbouring levels, as well as the upper clock state $|2,1\rangle$, which may enhance or reduce the effect.   
 
\begin{figure}[thb]
\centering
\includegraphics[scale=0.5]{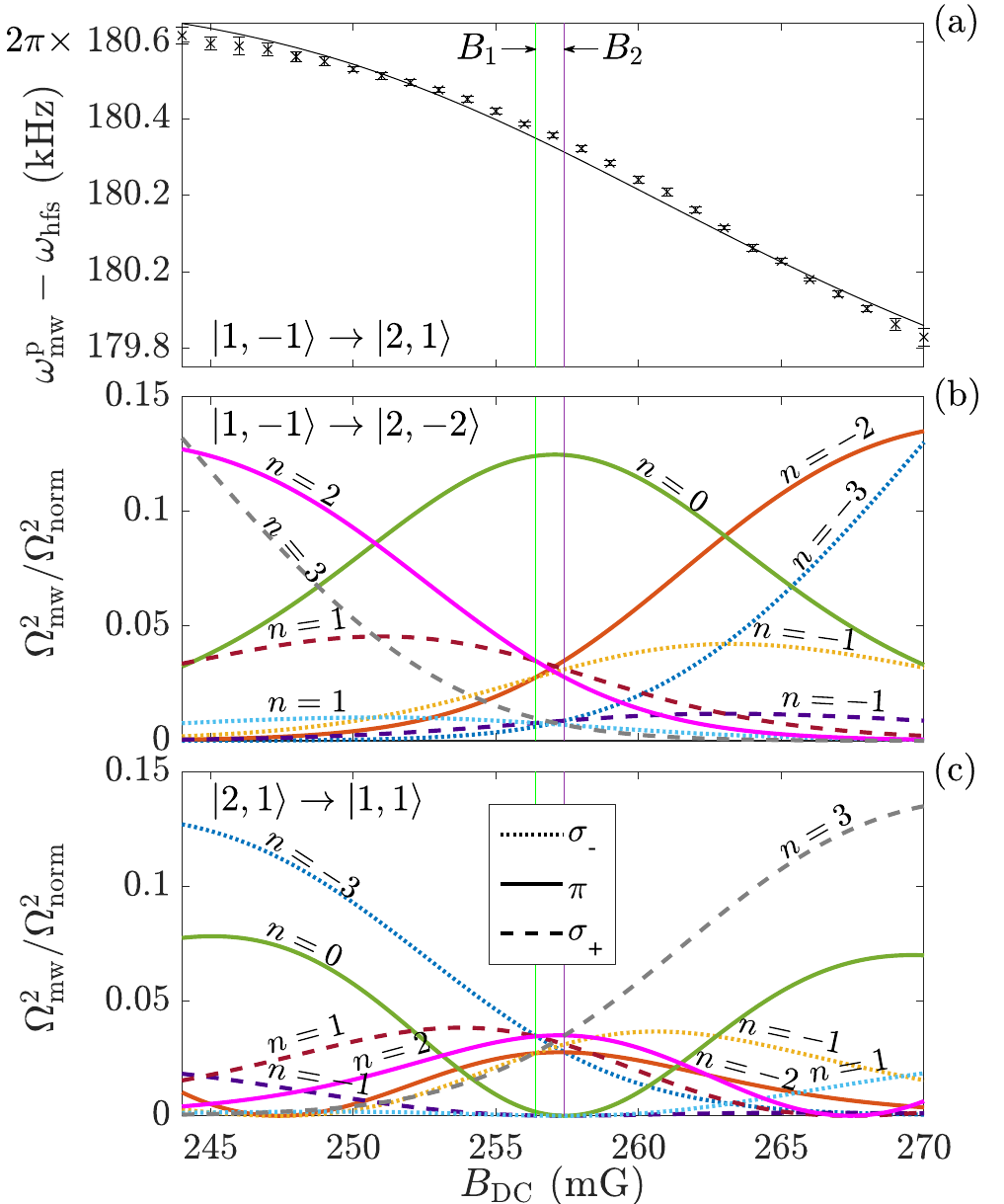}
    \caption{Properties of mw transitions involving the trappable, RF-dressed states $\ket{1,-1}$ and $\ket{2,1}$. The transition frequency between these two states, measured as probe detuning $\omega_\mathrm{mw}^\mathrm{p}-\omega_\text{hfs}$ is plotted in (a) as a function of static field $B_\mathrm{BC}$ across RF resonance. Experimental data points are shown together with a fit using Eqs.~\ref{Eq:Heff}-\ref{Eq:unshiftedEnergy} (solid line), with $B_\mathrm{RF}^\pm$ as free parameters. Without mw-dressing, a sloping behaviour arises from the difference in RF-resonant fields $B_F$, marked by vertical lines, locating the dressed potential minima.
    The lower panels show theoretical, field-dependent behaviour of relative coupling strengths $\Omega^2\sub{mw}/\Omega_\mathrm{norm}^2$ (Eqs.~\ref{eq:Rabi}-\ref{eq:Rabinorm})  across spectral groups ($n=-3$ to $3$) for the extremal transitions $\ket{1,-1}\to\ket{2,-2}$ and $\ket{2,1}\to\ket{1,1}$ in (b) and (c), respectively. Required mw polarisations are indicated by dotted, solid , and dashed lines for $\sigma_-, \pi, \sigma_+$, respectively. 
    Additional mw-dressing, red-detuned from a transition with rising slope in (b) or blue-detuned with falling slope in (c) can be used to counteract the field dependence shown in (a).}
\label{fig:4}
\end{figure}

\vspace{-2mm}
 
\section{Choice of Dressing Frequency}
\label{sec_dressfreq}

Figure~\ref{fig:4}(a) shows the weak static field dependence of the clock transition $\ket{1,-1}\leftrightarrow\ket{2,1}$ measured using group $n=1$, along with the theoretical prediction from Eq.~\ref{Eq:unshiftedEnergy}.
In the vicinity of the trapping minimum, i.e.\ near RF-resonance, the mw-transition frequency decreases with increasing static field. 

As any shift is approximately of the form $\delta E\approx \hbar\Omega^2/\Delta$, and $\Delta$ is an approximately even function, we require mw-dressing with a coupling strength behaviour of a form that opposes the field dependence seen in Fig.~\ref{fig:4}(a). The simplest first-order dressing choice is to pick one of the two clock states and manipulate this to match the field dependence of the other state. This dressing requires a transition which must either have a field dependence of $\Omega^2$ roughly matching or roughly opposite to that of the original behaviour. The sign of $\Delta$ must then be chosen appropriately such that the field dependence is reduced. As $\Omega\propto B\sub{mw}$, only the shape of the dependencies need to be considered, as the magnitude can be tuned via the mw-dressing power. 

In order to reduce the required dressing power, it is desirable to minimise the detuning from our chosen transition.
Due to our low RF-dressing power, transitions within each group are separated by only $12\times 2\pi$~kHz at the static field resonance. Without adiabatic ramping of field amplitudes, we find experimentally that a minimum detuning of $|\Delta|\geq10\times2\pi$~kHz is required in order to not risk undesired population transfer when mw-dressing with powers achievable in our experiment. We are therefore restricted to dressing on the outermost transitions from our chosen states in a given group, either red-detuned from  $\ket{1,-1}\to\ket{2,-2}$ or blue-detuned from $\ket{2,1}\to\ket{1,1}$. The coupling strengths for these two transitions when driven in different groups $n$ are plotted in Fig.~\ref{fig:4}(b) and (c) respectively.
Due to the difference in the sign of the required detuning we desire a coupling where $\Omega^2$ roughly matches the dependence of the clock transition if starting from $\ket{2,1}$, or shows opposite behaviour if starting from $\ket{1,-1}$.

From Fig.~\ref{fig:4} it can be seen that dressings for $n=-3,-2$ and $-1$ are of an appropriate form if manipulating $\ket{1,-1}$, or $n=1, -3$ for $\ket{2,1}$. In order to improve the cancellation beyond the first-order dependence using additional dressing fields, similar considerations can be used to guide the choice of further frequencies.
Note that in the case of higher static field and RF-dressing power, the groups and transitions therein will have a greater separation, so the limitations on the choice of transitions are alleviated, and a larger range of dressing frequencies becomes available.

\begin{figure}[thb]
\centering
\includegraphics[scale=0.5]{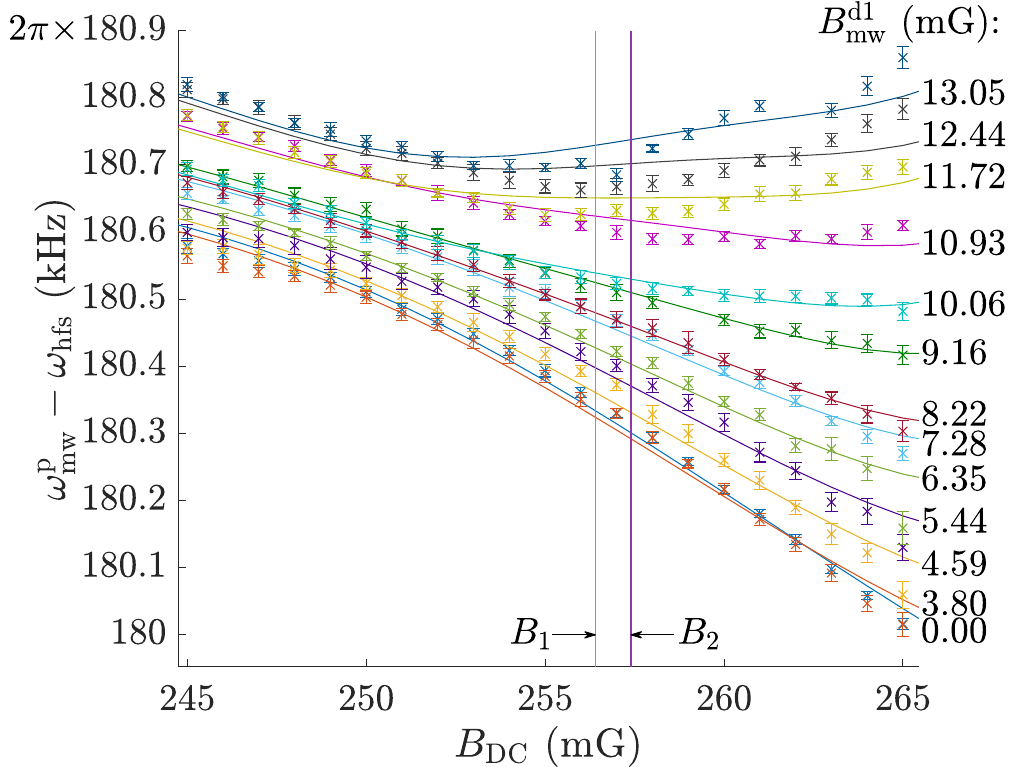}
    \caption{Static-field sensitivity of the transition frequency for $\ketto{1,-1}{2,1}$ in the presence of single-frequency mw dressing at $(\omega\sub{mw}^\mathrm{d1}-\omega\sub{hfs})/(2\pi)=-415$~kHz, i.e.\ red-detuned from group $n=-2$.
    Experimental data is shown along with the predicted transition frequency (solid lines) for a range of mw-dressing powers $B_\mathrm{mw}^\mathrm{d1}$. The model uses Eqs.~\ref{Eq:dressedmodel}-\ref{Eq:deltaE} with two free parameters $B_\mathrm{RF}^\pm$ for each curve. Two vertical lines mark the RF-resonant static fields $B_F$, locating the dressed potential minima at zero mw amplitude.
}
\label{fig:6}
\end{figure} 

\begin{figure*}[thb]
\centering
\includegraphics[scale=0.47]{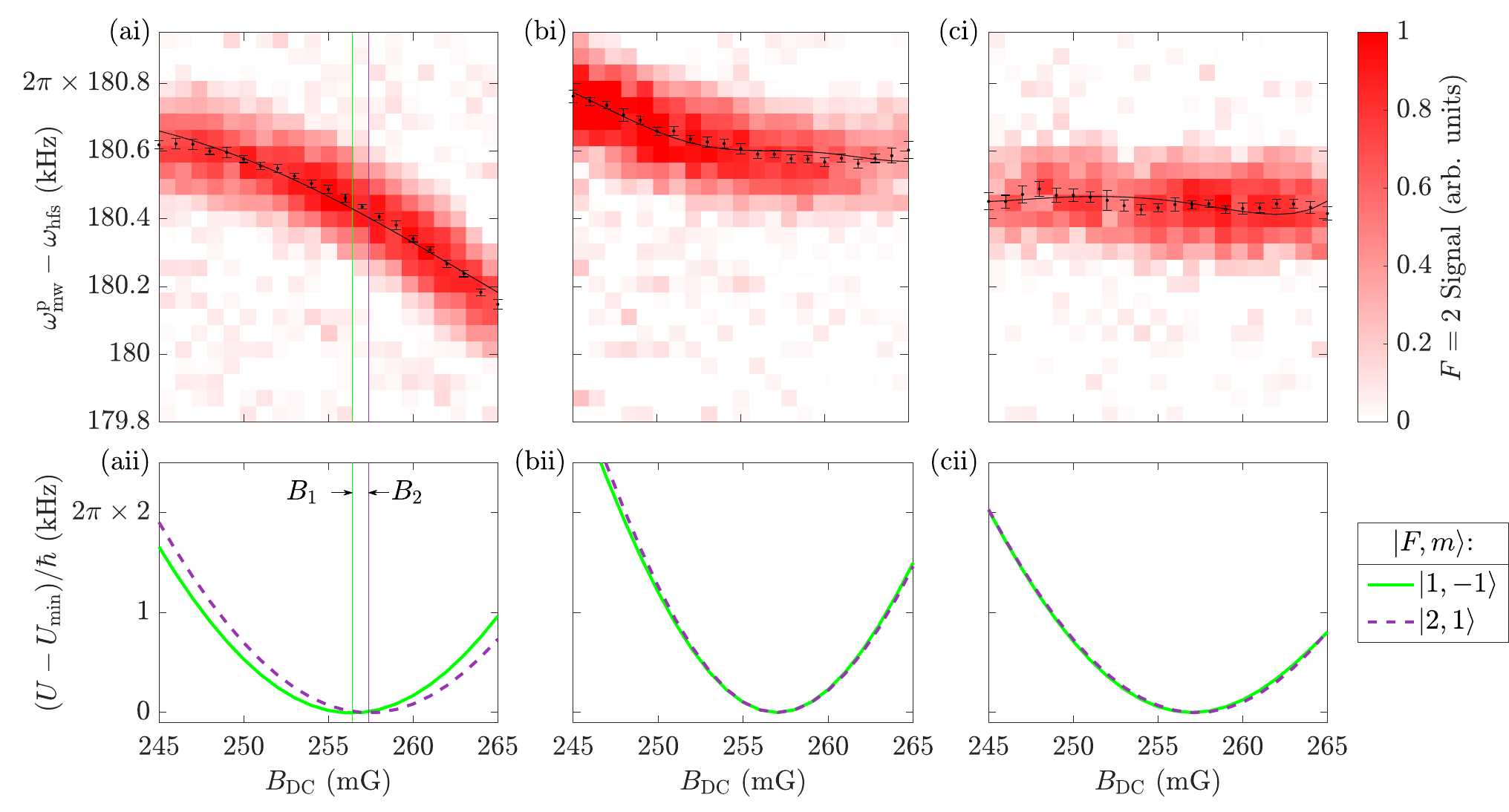}
    \caption{Higher-order compensation using multi-frequency dressing. The transition $\ketto{1,-1}{2,1}$ is probed by measuring population transferred to $F=2$ as a function of probe frequency $\omega_\mathrm{mw}^\mathrm{p}$ and static field $B_\mathrm{DC}$, represented as colourmaps in the top panels (i). In (a), the purely RF-dressed transition is shown, and in (b) and (c) additional single and double-frequency mw-dressing fields are applied. Experimentally determined centre frequencies are shown with model fits (solid lines) using $B_\mathrm{RF}^\pm$ as free parameters. In (c), the free parameters include a scale factor to account for mw-power compression (see main text). The corresponding RF-dressed potentials $U=E_{F,\bar{m}}$ inferred from these fits, offset by respective minima $U_\text{min}$, are shown in the corresponding bottom panels (ii) for $F=1$ (green solid lines) and $F=2$ (dashed purple lines). This analysis shows increased overlap of the two potentials, compensating the difference between original minima at $B_F$, marked by vertical lines in (a), as well as reduced shape mismatch.}
\label{fig:5}
\end{figure*} 

\vspace{-2mm}

\section{Multi-Frequency Dressing}

In the following we investigate the effect of using a single $\pi$-polarised dressing frequency at $\omega\sub{mw}^\mathrm{d1}-\omega\sub{hfs}=-415\times2\pi~\text{kHz}$ with amplitudes ranging over $B_\mathrm{mw}^\mathrm{d1} = 0-13$~mG, red-detuned $10\times2\pi$~kHz from the $n=2$ transition $\ket{1,-1}\to\ket{2,-2}$.
Atoms in a pure RF-dressed state $\ket{1,-1}$ are exposed for $3.5$~ms to a mw-probe with amplitude $B^\mathrm{p}\sub{mw}=1.5$~mG, where the amplitude of the coupling lin-perp component (driving $\sigma\pm$ transitions) is reduced to $0.38$~mG due to the orientation of the static field relative to the mw-antenna.
The measured transition frequencies as a function of static field are plotted in Fig.~\ref{fig:6} along with the theoretical fits using Eqs.~\ref{Eq:dressedmodel}-\ref{Eq:deltaE}. For each mw-power, we determine the value of two free parameters which are the amplitudes of effective $\sigma_\pm$-polarised RF-field components, as the model does not account for the back-action of the mw on the RF-resonance condition. A $2.5\%\pm0.1\%$ imbalance in amplitude between $\sigma_\pm$ components consistently found for all fits. We find that the effective RF-amplitudes are reduced proportional to the mw-dressing power, by a maximum of $30$~\% at the highest mw-power. The theoretical model shows good agreement with the data, especially at lower mw-dressing powers. At higher mw-powers the main approximations of the model become less valid, with small deviations of the observed field-dependence from the predictions, see the upper curves in Fig.~\ref{fig:6}.
It is clear that the sensitivity to static field is reduced with increased mw-power, with a vanishing slope at the RF-resonance for $B\sub{mw}^\mathrm{d1}\approx12$~mG.

In the top panels of Fig.~\ref{fig:5}, colourmaps of the final population fraction in $\ket{2,1}$ after mw probing are shown as a function of both static field and mw-probe frequency.  Similar plots were used to arrive at our measurements for transition frequency, using Gaussian fits, whose results are indicated with error bars. Fits to this data set, using Eq.~\ref{Eq:dressedmodel}-\ref{Eq:deltaE} are then used to extract values for the RF-dressed potential energies of the individual clock states. The corresponding results are shown in the lower panels. Panels (ai) and (aii) show the transition without mw-dressing and the corresponding potentials respectively. Near the potential minimum, we measure a field sensitivity of $\mathrm{d}\omega\sub{mw}/\mathrm{d}B\sub{DC}=(-29\pm2)\times2\pi$~Hz/mG, and a total change of $(474\pm11)\times2\pi$~Hz over the region $245\leq B\sub{DC}\leq265$~mG. In (b), the same plots are shown for a single mw-dressing at $\omega\sub{mw}^\mathrm{d1}-\omega\sub{hfs}=-415\times2\pi~\text{kHz}$ with amplitude $B\sub{mw}^\mathrm{d1}=12.4$~mG, which reduces the gradient near the potential minimum to $(-4\pm2)\times2\pi$~Hz/mG, and the total change over the measured region to $(199\pm14)\times2\pi$~Hz. In (bii), the two potentials show improved matching in field dependence over the undressed case.

We can now reduce the remaining field dependence further by adding another mw-frequency using the same method as outlined in Sec.~\ref{sec_dressfreq}. We chose a dressing frequency $10\times2\pi$~kHz blue-detuned from a transition $\ketto{1,1}{2,1}$ with a coupling strength roughly matching that of the residual field dependence. From Fig.~\ref{fig:4} we see that this transition in group $n=0$ is of the appropriate shape, hence our second dressing frequency is $\omega\sub{mw}^\mathrm{d2}-\omega\sub{hfs}=43\times2\pi~\text{kHz}$, which we find to be optimal at a nominal amplitude of $B\sub{mw}^\mathrm{d2}=9.1$~mG. In our fit we account for a mw-power reduction due to amplifier compression, as we cannot easily measure the power for both frequencies. Due to the additional free parameter, and the fact that the model breaks down at high mw-power, the fit should be treated as indicative only. Despite the inaccuracies of the model in this case, it is evident that the field dependence is drastically reduced. We see in Fig.~\ref{fig:5}(ci) that applying this in addition to our first-order dressing from (bi) does not markedly alter the gradient at resonance but does reduce the total change in field dependence across the measured static field range from $(199\pm14)\times2\pi$~Hz to $(72\pm18)\times2\pi$~Hz, which is comparable to our uncertainty on the measurement of the transition frequency. In (cii) we see that the mismatch between the two potentials has almost completely vanished. 
    
With two mw-dressing frequencies, we have shown that it is possible to cancel the field dependence of our chosen transition over the chosen static field range to less than the observed linewidth, which is limited by the time our free-falling atoms spend in the detection region. Our general method of identifying appropriate dressing frequencies may allow for extension to higher orders for increased coherence control. This model can easily be adapted to other atomic species and dressing situations.

\vspace{-2mm}

\section{Conclusion}

In this work, we have shown the potential for engineering transition frequencies to achieve increased coherence control using microwave-dressings in radio-frequency-dressed alkali atoms. The general method of identifying an appropriate dressing frequency, by considering the requirements on coupling coefficients and detuning, is outlined. The model allows for extension to multiple dressing frequencies, in order to generate a transition which is coherent to increasingly high order. We have demonstrated that it is possible with a single mw-dressing to fully cancel the first-order field dependence at the static field resonance, and have achieved frequency variation of less than the detection-limited linewidth over a wider static field range using two dressing frequencies. Improved detection resolution would enable investigation of how further dressing fields could extend the suppression of field sensitivity to higher orders. An improved model using Floquet analysis and the inclusion of second-order Zeeman effects arising from decoupling of nuclear and electronic spins at higher static fields may be required for other parameter ranges. The pair of coherently controlled, trappable synthetic clock states investigated here are promising for use in interferometry and quantum sensing.

\vspace{-5mm}

\section{Data availability}
The datasets generated for this paper are accessible
at \cite{multi-f-data} (Nottingham Research Data Management Repository).

\vspace{-5mm}

\section{Acknowledgments}
This research was funded by the  Engineering and Physical Sciences Research Council (Grants EP/M013294/1 and EP/Y005260/1). 
The authors would like to thank Konstantinos Poulios and Igor Lesanovsky for helpful discussions.

\bibliographystyle{apsrev4-1}
\bibliography{main}

\end{document}